\documentclass[sigconf,nonacm]{acmart}

\usepackage{multirow}
\usepackage{xurl}
\usepackage{pifont}
\AtBeginDocument{\providecommand{\checkmark}{\ding{51}}}

\newcommand{\yes}{\checkmark}
\newcommand{\no}{$\times$}
\newcommand{\circled}[1]{\raisebox{.5pt}{\textcircled{\raisebox{-.9pt}{\scriptsize #1}}}}
\newcommand{\sys}{\textsc{Cairn}}
\newcommand{\sllmv}{sllm+\vllm{}} 
\newcommand{\vllm}{vLLM}

\begin{document}

\title{Fast Serverless LLM Cold Starts by Sharing and Prewarming Startup Work}

\author{Hongjin Kim}
\affiliation{\institution{Korea Advanced Institute of Science \& Technology}\country{Republic of Korea}}
\author{Igjae Kim}
\affiliation{\institution{Korea Advanced Institute of Science \& Technology}\country{Republic of Korea}}
\author{Kwanghoon Choi}
\affiliation{\institution{Korea Advanced Institute of Science \& Technology}\country{Republic of Korea}}
\author{Minho Ha}
\affiliation{\institution{SK hynix}\country{Republic of Korea}}
\author{Jaehyuk Huh}
\authornote{Corresponding author.}
\email{jhhuh@kaist.ac.kr}
\affiliation{\institution{Korea Advanced Institute of Science \& Technology}\country{Republic of Korea}}

\renewcommand{\shortauthors}{Kim et al.}

\begin{abstract}
When a serverless LLM platform launches a model worker on demand, the cold start lasts roughly a minute, yet only 5--11\% of it is spent loading weights. Most of the remainder goes to compilation, memory profiling and runtime setup, which each server performs again even though the outcome is identical across servers or independent of the model. We present \sys{}. The server that is first to start a model publishes the weights and engine artifacts to a CXL memory pool shared across the cluster, and the remaining servers read them in place. Model-independent setup is completed ahead of time inside warm containers, and a template process replaces each container as soon as it is used. With ServerlessLLM on \vllm{} v0.20, six models from 1B to 14B, and two servers that share an emulated CXL pool, \sys{} speeds up cold starts by 4.1--5.2$\times$ on servers without a checkpoint of their own and reduces median TTFT on a request stream by 64\%.
\end{abstract}

\maketitle

\section{Introduction}
\label{sec:intro}

Serverless inference has become a common way to serve large language models: the platform starts a worker for a model only when the model has requests and stops it when the requests stop, so that many models share the same GPUs \cite{serverlessllm,deepserve,faascale}. The drawback is the \emph{cold start}. A request for a model with no running worker waits until one is started, and this wait is part of its time to first token (TTFT). For LLMs this wait is long, 57 to 85~s for models from 1B to 14B with ServerlessLLM on \vllm{} v0.20 (\sllmv{}), and hard to avoid, because a platform cannot keep a worker ready for every model (\S\ref{sec:motivation}).

Most prior work reduces the cold start by loading the model weights onto the GPU faster. Some systems make the transfer of weights to each server faster \cite{hydraserve,faascale}, and others prepare workers and weights in advance for models they predict will be needed \cite{warmserve,deepserve} (\S\ref{sec:prior}).

However, a cold start spends most of its time on work other than loading the weights, which takes only 5--11\%. Every new worker compiles the model, profiles GPU memory to size the KV cache, creates processes and imports a large serving stack, even though another server, or an earlier start, has already done the same work. Much of this work gives the same result on every server, or does not depend on the model at all. Yet every server repeats it on every start, and a burst that starts one model on $N$ servers repeats it $N$ times. Prior approaches either keep reusable state on individual servers or prepare it for specific models in advance, so they do not reuse model-specific startup artifacts across the cluster.

We classify cold-start work by two questions: does a step depend on the model, and is its result the same on every server? The answers give three classes, and Table~\ref{tab:classes} lists the techniques for the first two. \emph{Shareable} work encompasses model-specific operations whose results can be reused across compatible servers, including model loading, compilation, model inspection, and CUDA-graph memory estimation. \emph{Prewarmable} work does not depend on the model: creating processes, importing the serving stack and initializing CUDA. \emph{Process-bound} state, the CUDA graphs, belongs to the process that captures it. Shareable work needs to be done only once, and prewarmable work can be done before the model is known.

Cluster-wide shared memory makes sharing this work practical. A CXL memory pool lets every server of a cluster address the same memory in place, and their GPUs copy from it by DMA as from local host memory. One copy of a model's state in the pool can then serve every server, without a transfer per server.

We build \sys{} on these ideas. The first server to start a model publishes its weights, metadata, compile cache, compiler keys and CUDA-graph memory estimate to the pool as immutable entries, through a small coordinator, and every other server reads them in place, with no checkpoint of its own. Each server also keeps a \emph{warm container}, a front-end and an engine core that have finished every model-independent stage of startup and serve whichever model is deployed next. Because faster cold starts leave less time to replace a container that has been used, \sys{} forks replacements from a template process, a \emph{zygote}, instead of starting them from scratch.

On two servers sharing an emulated CXL pool, with \sllmv{} (\S\ref{sec:impl}), \sys{} cuts cold starts from 57--85~s to 11.0--21.0~s, 4.1--5.2$\times$ faster. Each of these cold starts runs on a server that has never served the model and holds no checkpoint of its own, reading only the entries that another server published. On a one-hour request stream that follows the Azure Functions trace, median TTFT falls by 64\%, from 32.5 to 11.8~s. In a simulated cluster of 16 servers, timeouts fall from 15.6\% to 2.1\% at 3.2 requests per second. We will release the source code of \sys{} upon publication.

To summarize, we make the following contributions:
\vspace{-8pt}
\begin{itemize}\setlength{\itemsep}{0pt}\setlength{\parskip}{0pt}
  \item We analyze the cold start of a current LLM engine and classify its work as shareable, prewarmable or process-bound.
  \item We design a CXL-emulated pool that lets every server reuse shareable artifacts that one server published.
  \item We design model-agnostic warm containers, replenished from a zygote, that do prewarmable setup before the model is known and can be applied regardless of the serving engine.
  \item We show that \sys{} makes cold starts 4.1--5.2$\times$ faster on servers that hold no checkpoint, and lowers TTFT and timeouts under load.
\end{itemize}

\section{Background and Related Work}
\label{sec:background}

\subsection{Serverless LLM Inference and Cold Starts}
\label{sec:anatomy}
\emph{Serverless computing} is a cloud model in which users deploy their code as functions and leave the provisioning and management of the servers that run it to the platform \cite{firecracker,sock}. The platform starts an instance of a function only when a request arrives, adds instances as the load grows, and removes them when the load falls, down to none when the function is idle. Users pay only while their instances run, and the platform shares its servers among the functions that currently have requests.

Beyond the flagship models of a few large providers, a growing number of LLMs are deployed by users themselves. They include open-source models used as they are, fine-tuned variants, and models trained for a particular domain or customer. Cloud platforms such as Amazon Bedrock \cite{bedrock}, Hugging Face Inference Endpoints \cite{hfendpoints} and Predibase \cite{predibase} now serve such models serverlessly. Users pay only while their model serves requests, and the platform can multiplex many models on the same GPUs.

Serverless inference is especially suited to models whose demand is low or hard to predict. Most models on such platforms are requested rarely, or in bursts separated by long idle periods \cite{serverlessllm}, as with a newly launched application or one used only at certain times. Keeping each of them loaded on its own GPUs would leave the most expensive and limited resource in the datacenter idle most of the time, and its cost would fall on users whether or not they send requests. A serverless LLM platform instead starts a \emph{worker}, an inference engine bound to one or more GPUs, only when a model has requests, and stops it when the requests stop. The same GPUs then serve many models over time, users pay only for the inference they use, and a burst of requests for one model is met by starting more workers for it \cite{serverlessllm,deepserve,faascale}.

The drawback of this approach is the \emph{cold start}. A request for a model with no running worker waits until one is started and initialized, and its time to first token (TTFT) includes this wait. For ordinary functions, platforms reduce cold starts by keeping an instance alive for a while after its last request (\emph{keep-alive}) and by preparing instances in advance. These techniques bring the cold start down to milliseconds or about a second \cite{sock,catalyzer,faasnap}. Both techniques are less effective for LLMs. Starting a worker loads gigabytes of weights onto a GPU and initializes an engine that compiles the model and profiles GPU memory, which takes tens of seconds rather than milliseconds. Keeping workers alive or prepared for many models, in turn, would occupy the GPUs that serverless inference is meant to share among models. For LLMs, cold starts are therefore both longer and harder to avoid.

\vllm{} \cite{vllm} is a widely used and rapidly developing open-source LLM inference engine. A \vllm{} worker runs as two processes: a \emph{front-end}, which receives and tokenizes requests, and an \emph{engine core}, which owns the GPU and runs the model. In a cold start, a newly created \vllm{} worker goes through six stages. In \emph{process start}, the front-end process is created and imports the Python libraries \vllm{} is built on, PyTorch, Transformers and \vllm{} itself; these libraries are large, and importing them takes seconds. In \emph{engine init}, the front-end launches the engine core as a separate process. The engine core imports the same libraries again, initializes CUDA, and sets up the executors that run the model. It also \emph{inspects the model}, reading its configuration to find which architecture to build. In \emph{model load}, the engine core builds the model's layers in GPU memory and copies the weights into them. The weights come from the model's \emph{checkpoint}, the files that store its weight tensors along with the configuration and tokenizer files needed to serve the model \cite{safetensors}. They are read from storage into host memory, through the page cache or pinned memory, and then copied to the GPU over PCIe.

The remaining stages prepare the model to run fast. In \emph{torch.compile}, \vllm{} compiles the model with PyTorch's compiler \cite{pytorch2}. The compiler first turns the model's Python code into a graph. It then generates GPU kernels from this graph, merging several operations into one kernel where it can. The compiled result is saved on disk in a \emph{compile cache}. The cache is looked up by a key computed from the installed software, so a later start on the same server can load the result instead of compiling again. In \emph{KV profile}, \vllm{} decides how much GPU memory to give the \emph{KV cache}. The KV cache holds the attention state of the requests being served. To size it, \vllm{} runs the model once on dummy input at the largest batch size, which shows how much memory a forward pass needs. It also captures a few CUDA graphs as a trial, to estimate how much memory the graphs will need. After setting aside memory for the weights, the forward pass and the CUDA graphs, \vllm{} gives all the remaining GPU memory to the KV cache. The last stage, \emph{CUDA graph}, records the GPU kernel launches of one decoding step so that they can later be replayed with little CPU time. \vllm{} records one CUDA graph for each of several batch sizes. However, a graph can only be replayed by the process that recorded it. The worker serves its first request only after this stage.

\subsection{CXL Memory Pools}
\label{sec:cxl}
Compute Express Link (CXL) lets a processor access memory attached to a device with ordinary loads and stores. Unlike disaggregation over a network, which copies remote bytes into a local buffer, CXL memory is part of the physical address space and is read in place, with latency within a small factor of local DRAM \cite{demystifyingcxl}. A CXL switch lets the servers of a cluster draw from a common pool of memory devices, which reduces memory stranded in individual servers \cite{pond,tpp}, and lets several of them map the same region.

Two properties of a CXL memory pool make it suited to model state. First, every attached server can read \emph{the same copy} of the data directly, without sending it over the network. Second, to a GPU it is \emph{host memory}: once mapped and registered, the GPU copies from it by DMA as from local pinned memory, with no intermediate copy. The difficulty is coherence: when several servers write the same region, their caches may disagree unless the hardware or a software protocol keeps them consistent. Model state is written once and only read afterwards, so it largely avoids this problem.

\subsection{Prior Work on LLM Cold Starts}
\label{sec:prior}
\textbf{Fast weight loading.} Most prior work on LLM cold starts reduces the time to load the weights onto the GPU. ServerlessLLM keeps checkpoints in a multi-tier store on each server and loads them with a pipelined loader \cite{serverlessllm}. HydraServe and FaaScale distribute a model across servers and overlap fetching with initialization \cite{hydraserve,faascale}. However, each server still receives and holds its own copy of the weights. Moreover, loading the weights accounts for only 5--11\% of a cold start (\S\ref{sec:breakdown}).

\textbf{Prewarming workers.} Prewarming systems prepare workers before requests arrive, so that a cold start skips part of its setup. DeepServe keeps model-agnostic pre-warmed workers, preloads predicted models into host memory, and sizes the KV cache offline \cite{deepserve}. WarmServe keeps a pool of worker processes with the serving libraries loaded, and loads the weights of models it predicts will be needed onto idle GPUs \cite{warmserve}. However, their model-specific preparation depends on predicting the next model and stays on particular GPUs or servers, and neither addresses how quickly a used worker is replaced. WarmServe is also built on \vllm{} 0.6.3, which by default does not use \texttt{torch.compile}, so its design does not address the longest stage of a cold start in the current \vllm{} (\S\ref{sec:breakdown}). It also runs the engine inside its API-server process and the GPU workers as Ray actors, so the processes it prepares in advance do not include the engine-core process that the current \vllm{} starts for each instance, which repeats the imports (\S\ref{sec:agnostic}).

\textbf{Reusing startup artifacts.} Another approach saves the results of startup work and reuses them at later starts. \vllm{} saves its compile cache on local disk, so a later start on the same server can skip compilation. Medusa materializes CUDA graphs offline so that capture need not be repeated at startup \cite{medusa}. However, these artifacts stay on the server that produced them, and model information, compiler keys and the CUDA-graph memory estimate are recomputed at every start. In a related analysis, Kabakibo et al.\ decompose the startup of \vllm{} on a single server and find that it is CPU-bound \cite{breakingtheice}.

\textbf{CXL memory pools.} Beluga stores the KV cache in a CXL memory pool that GPUs on several servers access directly through a CXL switch \cite{beluga}. These servers write to the pool concurrently, and the pool holds serving state rather than the state that a cold start produces. Table~\ref{tab:compare} summarizes what each approach targets in a cold start and whether its state is shared.

\begin{table}[t]
\caption{What prior systems and \sys{} target in a cold start, and how each loads the weights faster. Setup: model-agnostic setup; Shared: one copy of model state serves every server; Async load: the weight copy overlaps engine initialization; $^\dagger$only for models predicted in advance; $\triangle$: partly (KV-cache sizing only: computed analytically by HydraServe, offline by DeepServe).}
\label{tab:compare}
\vskip 0.1in
\begin{center}
\begin{footnotesize}\setlength{\tabcolsep}{1.6pt}
\begin{tabular*}{\columnwidth}{@{\extracolsep{\fill}}llcccc@{}}
\toprule
 &  & Compile, & & & Async \\
System & Weights & profile & Setup & Shared & load \\
\midrule
ServerlessLLM            & local cache            & \no         & \no   & \no  & \no \\
HydraServe               & parallel fetch         & $\triangle$ & \no   & \no  & \yes \\
WarmServe                & GPU preload$^\dagger$  & \no         & \yes  & \no  & \no \\
DeepServe                & DRAM preload$^\dagger$ & $\triangle$ & \yes  & \no  & \no \\
\midrule
\sys{}                   & shared pool            & \yes        & \yes  & \yes & \yes \\
\bottomrule
\end{tabular*}
\end{footnotesize}
\end{center}
\end{table}

\section{Motivation}
\label{sec:motivation}

\begin{figure}[t]
\centering
\includegraphics[width=\columnwidth]{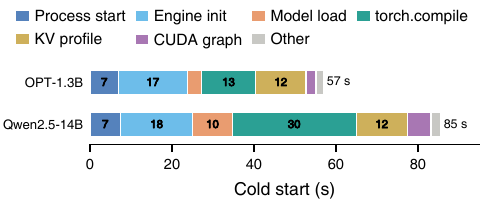}
\vspace{-12pt}
\caption{Cold start by stage for the smallest and largest of six models (\sllmv{}) on a server that has not served them before; checkpoint on NVMe, page caches cleared.}
\label{fig:breakdown}
\end{figure}

\subsection{Cold-Start Latency Breakdown}
\label{sec:breakdown}
Cold starts in serverless LLM inference are long, and they are hard to avoid (\S\ref{sec:anatomy}). Platforms cannot keep a worker warm for every model. Caching checkpoints in each server's DRAM does not solve this problem either. On a production trace, 20--46\% of ServerlessLLM's loads miss its host cache, most often when one model is scaled across several servers \cite{blitzscale}. Cold starts therefore remain common and arrive in bursts, and the length of each one matters.

Most of a cold start is not spent loading weights, so making the transfer faster, the focus of most prior work (\S\ref{sec:prior}), leaves most of the cold start in place. We measure the cold start of \sllmv{} on a server that has the serving stack installed but has not served the model before. The server has no copy of the checkpoint in host memory and no compile cache, and the checkpoint is on a local NVMe disk (\S\ref{sec:main}). It takes 57~s for OPT-1.3B to 85~s for Qwen2.5-14B (Figure~\ref{fig:breakdown}). Loading the weights is a small part: 6.9~s of 74.3~s for Qwen2.5-7B, against 24.4~s for compilation, 17.2~s for engine initialization, 12.6~s for KV-cache profiling and 7.5~s for process start. On our system, about 90\% or more of the cold start is spent outside weight loading.

\subsection{Repeated Model-Specific Work}
\label{sec:identical}
\label{sec:placement}
Every server that starts a model repeats work whose result is the same on every server. Most of a cold start produces artifacts that depend on the model but not on the server: compiled code, model information, Triton kernels \cite{triton}, compiler keys and the CUDA-graph memory estimate, besides the checkpoint itself. A server that has started a model before keeps some of them on its disk and in its memory, but a server starting it for the first time, the common case in serverless inference, produces them all again. We let two servers produce these artifacts independently, from empty caches, for the 3B, 7B and 14B models. Model information, compiler keys and the CUDA-graph memory estimate came out identical. The estimate even matched to the byte across five runs on each server. Compiled code and Triton kernels consist of the same files under the same paths, and differ only in a few bytes that record timing measurements. These artifacts are only megabytes in size, compared with gigabytes of weights, but producing them takes most of a cold start (Table~\ref{tab:artifacts}).

The reads are repeated too: each server holds its own checkpoint copy in pinned memory, and Triton's compiled library (396~MiB) is read in full on every cold start only to compute a cache key. A burst that starts one model on $N$ servers therefore reads, loads and compiles it $N$ times. The reusable state also stays only on the server that made it, which limits the scheduler's choices. A worker starts quickly only on a server that already holds the model's reusable state, and the workers already running on that server may have to be moved to another server to make room.

\begin{table}[t]
\caption{Model-specific artifacts of Qwen2.5-7B: their size and the cold-start time each saves when it alone is reused.}
\label{tab:artifacts}
\vskip 0.1in
\begin{center}
\begin{footnotesize}\setlength{\tabcolsep}{4pt}
\begin{tabular*}{\columnwidth}{@{\extracolsep{\fill}}lrr@{}}
\toprule
Artifact & Size & Saves \\
\midrule
compile cache and Triton kernels & 34.0~MB & 27.2~s \\
model information                & 748~B   & 10.0~s \\
compiler keys                    & 2.1~KB  & 2.9~s \\
CUDA-graph memory estimate       & 402~B   & 3.6~s \\
metadata                         & 11.5~MB & -- \\
\bottomrule
\end{tabular*}
\end{footnotesize}
\end{center}
\end{table}

\subsection{Repeated Model-Agnostic Setup}
\label{sec:agnostic}
\begin{figure}[t]
\centering
\includegraphics[width=\columnwidth]{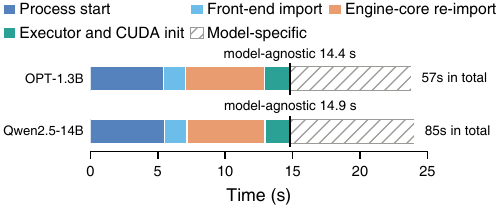}
\vspace{-12pt}
\caption{Model-agnostic part of the cold start for the smallest and largest of six models with \sllmv{}: the first 24~s of cold starts that take 57~s and 85~s in total (Figure~\ref{fig:breakdown}). Process start includes the Ray actor and the PyTorch import. The hatched model-specific part is model inspection, shown for reference.}
\label{fig:agnostic}
\end{figure}

Besides model-specific work, every cold start pays for setup that does not depend on the model, such as creating processes, importing the serving stack, starting executors and initializing CUDA. None of it needs to know which model will be served, yet a worker cannot do it until a request names the model, so it lies on the critical path of every cold start. Figure~\ref{fig:agnostic} shows it for the smallest and the largest model. Apart from about 9~s of model inspection, process start and engine initialization are this kind of setup. They take 14.4~s for OPT-1.3B and 14.9~s for Qwen2.5-14B, and differ by at most 0.5~s across all six models. Much of it is even done twice. The front-end imports PyTorch and \vllm{}, and the engine core then imports them again. This is because \vllm{} starts the engine core with \texttt{spawn}, which creates a fresh Python process. \vllm{} uses \texttt{spawn} when it starts the main process itself, as with \texttt{vllm serve}, and must use it inside a Ray actor \cite{ray} or once CUDA is initialized. The second round of imports alone costs 5.8--5.9~s. A process that has done this setup before any model is known, a \emph{warm container}, could take on the model when a request arrives and skip it.

\label{sec:refill}A warm container, however, helps only if one is ready at every cold start. Warm pools prepare processes in this way for serverless functions \cite{sock}, and each cold start consumes one prepared process. A replacement must therefore be ready before the next request for that GPU, and the faster cold starts become, the less time a replacement has. Replacing it by starting new processes repeats the imports it was meant to save, and takes 13.9~s. Once model-specific work is shared and setup is prepared, a cold start itself takes only 11.0--21.0~s (\S\ref{sec:main}). For small models this is less than the replacement time, for example 11.0~s for OPT-1.3B. The next request can then arrive before a prepared process is ready, and spawning the replacement earlier has its own cost.

\begin{figure}[t]
\centering
\includegraphics[width=\columnwidth]{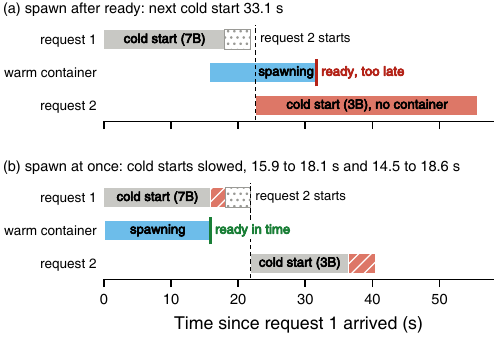}
\vspace{-12pt}
\caption{Replacing a warm container by spawning: (a) after the previous instance is ready, (b) at once.}
\label{fig:refill}
\end{figure}

Figure~\ref{fig:refill} shows both choices, measured with \sllmv{} and all mechanisms. In this experiment, request~2 asks for a different model (Qwen2.5-3B after 7B) and arrives as soon as request~1 finishes. We ran this pair 10 times, and each bar shows the median. In (a), the replacement is spawned once request~1's instance is ready, so as not to slow its cold start. It was still being built when request~1 released the GPU (dotted) in all 10 pairs, and request~2 paid an ordinary cold start of 33.1~s. In (b), the replacement is spawned at once. It was ready about 6~s before request~1 released the GPU in all 10 pairs. However, it competed for the CPU with the cold starts it overlapped. These took 18.1 and 18.6~s, compared with 15.9 and 14.5~s without contention (hatched).

\section{Design}
\label{sec:design}

\subsection{Overview}
\sys{} removes repeated work in two ways, summarized in \S\ref{sec:summary} and Table~\ref{tab:classes}. Work that is the same on every server is done once and \emph{shared} through a cluster-wide pool (\S\ref{sec:pool}). Setup that does not depend on the model is done before the model is known, in a \emph{warm container} on each server. Because each cold start consumes one warm container, replacements are \emph{forked} from a template process rather than built from scratch (\S\ref{sec:shell}). What remains is CUDA graph capture, whose graphs are bound to the process that captures them, so \sys{} performs it as usual. Restoring captured graphs is the subject of Medusa \cite{medusa} and is orthogonal to \sys{}.\label{sec:classiii}

\begin{figure}[t]
\centering
\includegraphics[width=\columnwidth]{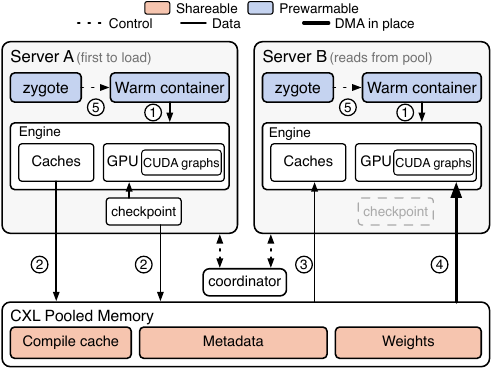}
\vspace{-11pt}
\caption{Two servers sharing a pool in \sys{}. Numbers follow the flow described in \S\ref{sec:design}; colors follow the classes of Table~\ref{tab:classes}.}
\label{fig:overview}
\end{figure}

Figure~\ref{fig:overview} shows the resulting flow on two servers. A \emph{deployment}, a request to start serving a model on a GPU, is handed to a waiting warm container, which becomes the engine (\circled{1}). The first server to load a model, here server~A, produces its artifacts on the ordinary path and writes them through to the pool (\circled{2}). Any other server, here server~B, then restores the model's metadata, compile caches, compiler keys and CUDA-graph memory estimate from the pool (\circled{3}). Its GPU copies the weights directly from the pool (\circled{4}). Compilation, profiling and graph capture then run with most of their work already done. Meanwhile the zygote forks the next warm container (\circled{5}). Whenever a container or a pool entry is missing, the server takes the ordinary path, so neither mechanism can make a cold start slower than it would otherwise be.

\subsection{Sharing Model-Specific Artifacts}
\label{sec:pool}
\label{sec:readpath}
\label{sec:layout}
\label{sec:artifacts}
\label{sec:poolfit}
To be reused across servers, the artifacts of \S\ref{sec:identical} must be stored where every server can access them. Prior systems keep them where only one server can use them (Table~\ref{tab:compare}). They store the artifacts in a server's DRAM or storage, transfer a copy to each destination, or load weights into idle GPU memory by prediction. A cluster-shared memory pool instead holds one copy that every server's GPU reads in place, at close to local-memory bandwidth, with capacity sized for the cluster. Model state is also well suited to shared memory because, unlike the KV cache \cite{beluga}, it is written once and only read afterwards.

We store every shareable artifact of a cold start in the pool as an \emph{entry}. The entries of a model are its weights, stored as one contiguous blob, its metadata, and its compile cache. The compile-cache entry also holds the model information and the Triton kernels. The compiler keys of the software environment and the CUDA-graph memory estimate are stored as entries as well. Each entry is keyed by the model and a \emph{compatibility fingerprint}. The fingerprint covers the GPU type, the versions of the engine, compiler and libraries, and the engine options that change compiled code or graph capture. An artifact made under a different fingerprint is never matched, and a miss only means that the artifact is produced on the ordinary path. Compiler keys depend only on the installed software, so one entry serves every model.

To publish and locate these entries, a small \emph{coordinator} owns the pool's index of entries and free space. A server that needs an entry asks the coordinator. If the entry is valid, the coordinator returns its location. If no server is producing it, the coordinator assigns the entry to the server and allocates a free range for it. The server then produces the artifact on the ordinary path and writes it to that range, writing weights through as they are read from storage. If another server is already producing the entry, the server waits until it becomes valid. A burst of cold starts for one model thus produces each entry once.

Since the pool has no hardware cache coherence across servers, publication follows two rules. Each range has a single writer and multiple readers (SWMR), and each entry is immutable once it becomes valid. The coordinator marks an entry valid only after the writer flushes the range from its CPU caches and reports completion. Readers invalidate their cached lines of the range before reading it. If the writer fails, the entry never becomes valid, and the coordinator reassigns it after a timeout. Servers therefore need neither atomic operations nor coherence on the pool.

Once an entry is valid, a server uses it according to its type. The GPU reads the weights in place, copying them directly from the pool with no copy in pinned memory and no local checkpoint. The compile cache and model information are restored into the engine's own caches before the engine starts, so the engine finds a warm cache and skips the work that produced it. The compiler keys and the CUDA-graph memory estimate are instead \emph{injected} into the engine, which uses them rather than computing them. In addition, we load the weights \emph{asynchronously}. Compilation, profiling and graph capture need the parameters' addresses and shapes but not their values, so the copy overlaps with initialization and is confirmed complete before the first request. This also applies to weights read from storage.

Because every server reads the same entries, a model's state no longer belongs to the server that produced it. A worker starts as quickly on any server in the cluster as on the server that published the model's entries. Placement can therefore follow load rather than state, and no other workers have to be moved to another server to make room.

\subsection{Prewarming Model-Agnostic Setup}
\label{sec:shell}
\begin{figure}[t]
\centering
\includegraphics[width=\columnwidth]{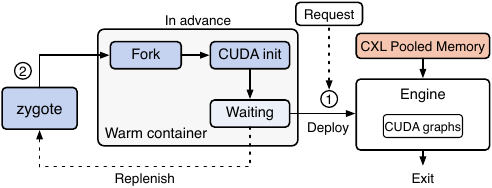}
\vspace{-11pt}
\caption{Life of a warm container.}
\label{fig:warm}

\end{figure}

Figure~\ref{fig:warm} shows the life of a warm container. A warm container is a pair of processes, a front-end and an engine core, that has completed all the model-independent setup of \S\ref{sec:agnostic} before any model is known. Both processes have imported the serving stack, and the engine core has initialized CUDA and created its CUDA context. Since the engine core is created in advance, the second round of imports in a newly spawned engine core is avoided. The container then waits without a model. Configuration, tokenizer, model construction, weights, compilation, profiling and graph capture are left for the deployment.

When a deployment request for model $m$ arrives, we assign the waiting container to it (\circled{1}). The front-end runs the engine's usual startup with the deployment's arguments, except that it uses the waiting engine core instead of creating a new one. Specifically, the engine core continues from configuration and model loading, restores $m$'s artifacts from the pool, loads the weights, and then compiles, profiles and captures graphs as an ordinary engine would. As a result, the engine and the serving system above it see an ordinary start. The container serves $m$ and exits when the deployment ends. We do not reuse it for another model, since its engine already holds the distributed and compilation state of $m$. Because a warm container holds no model state, a single container per GPU can serve any model without prediction.

However, each deployment consumes the container on its GPU, and a replacement must be ready before the next deployment arrives (\circled{2}). Building the replacement from new processes repeats the imports that the container is meant to save, and takes 13.9~s (\S\ref{sec:refill}). Copying a waiting container with \texttt{fork} is also not feasible, because its engine core has initialized CUDA. \texttt{fork} copies the memory of a process, but not the state that the GPU driver keeps for that process. Once CUDA is initialized, the driver maintains a CUDA context for the process, which maps GPU memory into its address space and relies on threads created by the driver. A child created by \texttt{fork} inherits the address mappings but neither these threads nor the context itself, and therefore cannot use CUDA. For this reason, CUDA does not support \texttt{fork} after initialization, and both PyTorch and \vllm{} start a new process with \texttt{spawn} in this case.

\label{sec:zygote}To address this, we fork replacements from a \emph{zygote}. A zygote is a template process that has imported the serving stack but has never initialized CUDA, a technique also used by Android and SOCK to start applications and serverless functions quickly \cite{sock}. Each server keeps one zygote. The engine core of a replacement is forked from the zygote, inherits its imports, and then creates its own CUDA context (\circled{5} in Figure~\ref{fig:overview}), while its front-end starts as a new process. Since forking requires little CPU, we start a replacement as soon as a container is taken, without slowing down the cold start it overlaps. If the zygote cannot produce a child that initializes CUDA, we spawn containers instead. In addition, no failure blocks a deployment. Without a ready container, or if the handoff fails, the deployment follows the ordinary path, and failed containers are replaced with back-off.

\subsection{Summary}
\label{sec:summary}
Table~\ref{tab:classes} lists the techniques of \sys{} by the class of work they target. For shareable work, a shared pool holds one copy of each model's state for the cluster. Weights are written through to the pool on the first load and loaded from it asynchronously. The compile cache and model information are restored from the pool. The compiler keys and the CUDA-graph memory estimate are injected into the engine instead of being computed. For prewarmable work, a warm container on each GPU has finished every model-independent step before a deployment arrives, and the zygote replaces it as soon as it is taken. Process-bound state, the CUDA graphs, is captured as usual. Each technique falls back to the ordinary path when its state is missing, so they can be enabled one at a time (\S\ref{sec:ablationstudy}).

\textbf{Applicability to other engines.} The cold starts of other LLM serving engines go through stages similar to those of \vllm{}, so warm containers can be applied to them as well. SGLang \cite{sglang} and TensorRT-LLM \cite{trtllm} also start each GPU worker as a new process, SGLang with \texttt{spawn} and TensorRT-LLM through MPI even on a single GPU. Each worker then imports the engine again, sets up its worker objects, and initializes CUDA and its communication library. These stages correspond to process start, the engine-core re-import, and executor and CUDA initialization in \vllm{} (Figure~\ref{fig:agnostic}). Measured with the same stages, they take 22.3~s in SGLang and 32.5~s in TensorRT-LLM for Qwen2.5-7B, and they change by less than 1~s across models. A warm container can therefore prepare them in advance in either engine.

The pool applies to these engines as well. Both SGLang and TensorRT-LLM support \texttt{torch.compile} as an option that is off by default, and when it is enabled, a server can restore its compiled code from the pool as in \vllm{}. SGLang also compiles some kernels with nvcc at run time and keeps them in a cache keyed by kernel shape. Without this cache, a cold start of Qwen2.5-7B takes 62--70~s longer, and the cache can be reused across models whose kernels have the same shapes.

\begin{table}[t]
\caption{Classes of cold-start work and the techniques \sys{} uses for the shareable and prewarmable classes.}
\label{tab:classes}
\vskip 0.1in
\begin{center}
\begin{footnotesize}\setlength{\tabcolsep}{3pt}
\setlength{\aboverulesep}{2.5pt}\setlength{\belowrulesep}{2.5pt}
\newcommand{\dblrule}{\specialrule{0.4pt}{1pt}{1pt}\specialrule{0.4pt}{0pt}{2.5pt}}
\begin{tabular}{@{}p{0.47\columnwidth}l@{}}
\toprule
Target & \sys{} \\
\midrule
\multicolumn{2}{@{}l}{\textbf{Shareable}} \\
\dblrule
\multirow{3}{*}{Weights}   & Shared pool \\
                           & Weight write-through \\
                           & Asynchronous weight load \\
\cmidrule{1-2}
\begin{tabular}[c]{@{}l@{}}Metadata / Compiled code /\\ Model inspection\end{tabular} & Shared pool \\
\cmidrule{1-2}
Compiler keys              & Compiler-key injection \\
\cmidrule{1-2}
CUDA-graph memory estimate & \begin{tabular}[c]{@{}l@{}}CUDA-graph memory\\ \quad estimate injection\end{tabular} \\
\midrule
\multicolumn{2}{@{}l}{\textbf{Prewarmable}} \\
\dblrule
\begin{tabular}[c]{@{}l@{}}Processes / Imports /\\ CUDA initialization\end{tabular} & Warm container \\
\cmidrule{1-2}
Replacing a used container & Zygote replenishment \\
\bottomrule
\end{tabular}
\end{footnotesize}
\end{center}
\end{table}

\section{Implementation}
\label{sec:impl}

\sys{} is implemented on a fork of ServerlessLLM \cite{serverlessllm}. ServerlessLLM serves models with a Transformers backend by default. It also provides a \vllm{} backend, but only for an earlier version of \vllm{} and with limited functionality. We ported that backend to \vllm{} v0.20 and its V1 engine (677 lines), including the integration of ServerlessLLM's checkpoint store with the new model loader. We use the result, which we call \sllmv{}, throughout. On top of this port, \sys{} adds 4{,}435 lines of code. The pool store and its tools take 1{,}247 lines of Python, and the pool read path, write-through and asynchronous load take 1{,}626 lines of C++ in \texttt{sllm-store}. The warm container, zygote and container manager take another 1{,}082 lines of Python, and a \vllm{} plugin takes the remaining 480.

\textbf{Pool.} Lacking a CXL switch, we emulate two servers sharing a pool with two virtual machines on one host, using remote NUMA memory as a stand-in for CXL memory as in prior work \cite{pond,tpp}. A memory region on the host's remote NUMA node is attached to both VMs as the same shared virtual NVDIMM. Each guest maps it as a \texttt{devdax} device, which exposes memory without a file system. The guest registers it for DMA and reads it in place, as it would a CXL pool. The coordinator's index is a slot table at the head of this region, updated with compare-and-swap under leases that writers renew. The store maps an entry with \texttt{mmap(MAP\_POPULATE)}, registers it with \texttt{cudaHostRegister} and issues one asynchronous copy per tensor; write-through fills the pool and the GPU from the same direct-I/O buffer.

\textbf{Engine hooks and zygote.} Compiler keys and CUDA-graph memory estimates are injected through a \vllm{} plugin. The warm container replaces the function from which \vllm{} V1 obtains its engine-core process factory, so that \vllm{} receives the waiting child as if it had just started it. In the zygote, libraries that query the GPU at import time are given answers recorded in advance, and those that need a CUDA context are imported in each child after it creates one.

\section{Evaluation}
\label{sec:eval}

\subsection{Methodology}
\label{sec:setup}
\textbf{Environment.} We use two servers emulated on one host (\S\ref{sec:impl}). Each server has its own kernel, an NVIDIA RTX A6000 (48~GB), 8 dedicated cores, 64~GB of memory and its own disk. The two servers share only a 96~GB pool, which the GPU reads at 22.4--22.7~GB/s, compared with 25.3~GB/s from local pinned memory. Each server's disk is an image file on a host NVMe drive, which the server reads at 4.9~GB/s. We use six models with weights of 2.47 to 29.54~GB, namely Llama-3.2-1B-Instruct \cite{llama3}, OPT-1.3B \cite{opt}, and Qwen2.5-1.5B, 3B, 7B and 14B-Instruct \cite{qwen25}.

\textbf{Comparisons.} All experiments use \sllmv{}, ServerlessLLM with its \vllm{} backend ported to \vllm{} v0.20 (\S\ref{sec:impl}). It starts each instance inside a Ray actor, reads weights through the ServerlessLLM store, and starts the engine core with \texttt{spawn}. The single cold-start experiments run in four conditions. Besides \emph{none}, which uses no mechanism, \emph{warm container} uses warm containers that the zygote replaces as soon as one is taken (\S\ref{sec:shell}), and \emph{pool} keeps weights, compile caches, compiler keys, and CUDA-graph memory estimates in the pool. \emph{All} combines the two and all the other methods we used. On the request stream, we also compare two ways to replace a warm container: spawning a new one and forking it from the zygote. In every condition, ServerlessLLM's router is notified of instance readiness by an event rather than by its one-second polling, so the baseline does not pay for polling.

\textbf{Benchmarks.} We use two workloads. The first is a single cold start: one request for a model that the server has not served before, generating 8 tokens greedily from a fixed 16-token prompt (\S\ref{sec:main}--\S\ref{sec:ablationstudy}). The second is a request stream. Its arrivals follow the Azure Functions trace \cite{azurefunctions}, with functions mapped round-robin to models, rates scaled to a target load and each model's arrivals drawn from a Gamma process with a given burstiness (CV) \cite{alpaserve}. Each request takes its prompt and output length from the first turn of a ShareGPT conversation \cite{sharegpt}. We run this workload for one hour on our two servers at CV~1 (\S\ref{sec:overlapping}), and in a simulated cluster while varying the request rate and the CV (\S\ref{sec:scale}).

\textbf{Metrics.} For a single cold start, we report its time to first token (TTFT), from the request's arrival to its first token, split into the stages of \S\ref{sec:anatomy}. For the request stream, we report the median and tail of TTFT and the share of cold starts that found a warm container. In the simulation, we also report the share of requests that time out, and count each of them as 300~s.

\subsection{End-to-End Cold Start}
\label{sec:main}
\begin{figure}[t]
\centering
\includegraphics[width=\columnwidth]{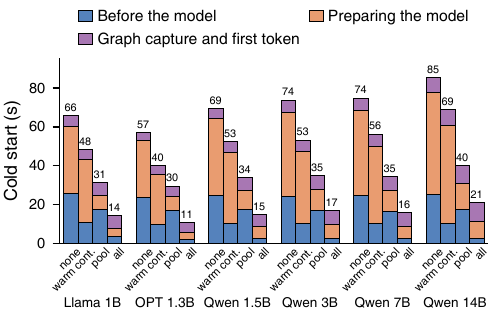}
\vspace{-12pt}
\caption{Cold start with \sllmv{} for six models in four conditions, in three parts in the order they run: before the model is known (process start and engine initialization), preparing the model (loading, compilation and KV profiling), and graph capture and the first token.}
\label{tab:main}
\end{figure}

Each cold start runs on a server that has the serving stack installed but has not served the model. Before each measurement, we drop the model's files from the page caches of the VM and the host. We also delete the server's local compile cache. When weights come from disk, we restart the store so that it holds no DRAM copy. The pool entries stay in place and were published by another server. The serving libraries are imported once beforehand as on any server that has run the stack before. The instance is released when the request finishes. Sections~\ref{sec:ablation} and~\ref{sec:ablationstudy} use the same procedure.

Figure~\ref{tab:main} shows the cold start of the six models in the four conditions with \sllmv{}, split into three parts in the order they run. A warm container removes most of the first part, which runs before the model is known, and the pool shrinks the second, which prepares the model. The last part, graph capture and the first token, remains. With both mechanisms, \sys{} achieves 4.1--5.2$\times$ faster cold starts than \sllmv{}, at 11.0--21.0~s. The baseline grows with model size, from 57 to 85~s. With all mechanisms, the cold start grows much less. Even compared with a server that has served the model before and keeps its local compile cache and ServerlessLLM's DRAM copy, \sys{} is about 2$\times$ faster for the 3B to 14B models on a server that has never served it.

\subsection{Breakdown by Mechanism}
\label{sec:ablation}
\begin{figure}[t]
\centering
\includegraphics[width=\columnwidth]{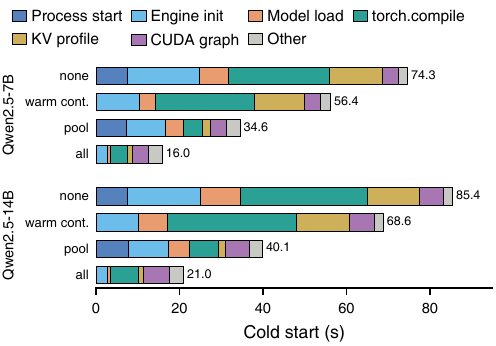}
\vspace{-12pt}
\caption{Cold start with \sllmv{} by stage for Qwen2.5-7B and 14B with each mechanism on or off.}
\label{fig:e2e}
\end{figure}

Figure~\ref{fig:e2e} breaks the cold start of Qwen2.5-7B and 14B into stages, with each mechanism on or off. The two mechanisms remove different stages, so their effects add. For the 7B model, the warm container alone reduces the cold start by 24\%, the pool alone by 53\%, and both together by 78\%. The warm container reduces process start by 99\% and engine initialization by 41\%. The rest is mostly model inspection, which the pool removes by restoring the model information. The pool reduces compilation by 81\%, KV-cache profiling by 86\%, and engine initialization by 46\%. On NVMe, the weight path itself saves little time (6.9 to 4.4~s). With both, process start and engine initialization fall by 97\% and 85\%.

With all mechanisms, 16.0~s remain for the 7B model. Loading the cached compiled graph takes 4.1~s and engine initialization 2.5~s. CUDA graph capture, which is process-bound, takes 3.7~s, and request routing, server readiness and the first token take another 3.6~s.

\subsection{Ablation Study}
\label{sec:ablationstudy}
\begin{figure}[t]
\centering
\includegraphics[width=\columnwidth]{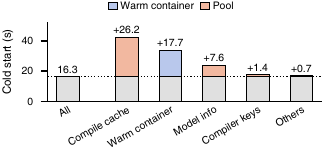}
\vspace{-13pt}
\caption{Cold start of Qwen2.5-7B with \sllmv{}, with all mechanisms and with one component removed. Each bar shows the increase over all mechanisms, colored by the mechanism the component belongs to.}
\label{fig:ablation}
\end{figure}

To understand how much each component contributes when all the others are enabled, we remove one component at a time from the full system. Figure~\ref{fig:ablation} shows the resulting cold start of Qwen2.5-7B. The full system takes 16.3~s in this experiment. Every component still matters with all the others on, and the compile cache matters most. Without it, the cold start becomes 2.6$\times$ as long (42.5~s), because Triton kernels are compiled again during the first profiling pass, which also lengthens KV profiling. Removing the warm container lengthens the cold start by 109\%, removing model information by 47\%, and removing the compiler keys by 9\%. The CUDA-graph memory estimate, asynchronous load and pooled weights add 0.7~s together, shown as \emph{Others} in Figure~\ref{fig:ablation}. The last two are small because our baseline already holds a checkpoint copy on every server's NVMe: the 7B checkpoint loads from it in 3.1~s, and asynchronous load hides that behind compilation. On this storage, the benefit is the memory and storage a server no longer needs. With the pool, a server keeps only 11.5~MB of metadata on its disk for the 7B model, instead of a 15.2~GB checkpoint. We also turned on each artifact alone, starting from a cold start without any mechanism (72.9~s). The compile cache then saves 37\%, model information 14\%, and the compiler keys and the CUDA-graph memory estimate 4\% and 5\%. Together, the compile cache and model information save 48\% (Table~\ref{tab:artifacts}).

\subsection{Replenishment Under a Request Stream}
\label{sec:overlapping}
\begin{figure}[t]
\centering
\includegraphics[width=\columnwidth]{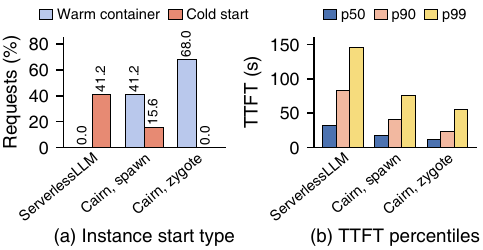}
\vspace{-12pt}
\caption{A one-hour request stream at CV~1 on two servers, for \sllmv{} and for \sys{} with spawned or zygote replacements. (a) Share of the 403 requests that started an instance with a warm container (Warm container) or without one (Cold start); the rest found a running instance. (b) TTFT percentiles.}
\label{fig:replenish}
\end{figure}

To evaluate whether warm containers are replaced fast enough to be ready at each cold start, we run the request stream with two ways to replace a used container, spawning new processes or forking from the zygote. Forking prepares a replacement in 7.6~s, against 13.9~s for spawning. The stream sends 403 requests over one hour to the five models other than Qwen2.5-14B, at CV~1 and a target load of 0.05 requests per second per GPU. Each request uses a ShareGPT prompt and its answer length. Every run starts with empty caches and an empty pool, and an instance is released as soon as it has no pending request. We choose CV~1 because cold starts on a server then often follow each other faster than a spawned replacement can be prepared.

Figure~\ref{fig:replenish} shows the results. With zygote replacements, every cold start finds a warm container. With spawned replacements, 28\% of cold starts find none and fall back to the ordinary path (Figure~\ref{fig:replenish}a). \sys{} starts more instances than \sllmv{}, because with keep-alive zero its shorter cold starts leave fewer requests waiting for an instance that is still starting. Compared with \sllmv{}, zygote replacements reduce median TTFT by 64\%, to 11.8~s, and the 90th percentile by 72\%, against 44\% and 51\% with spawned replacements (Figure~\ref{fig:replenish}b).

\subsection{Simulation at Cluster Scale}
\label{sec:scale}
\begin{figure}[t]
\centering
\includegraphics[width=\columnwidth]{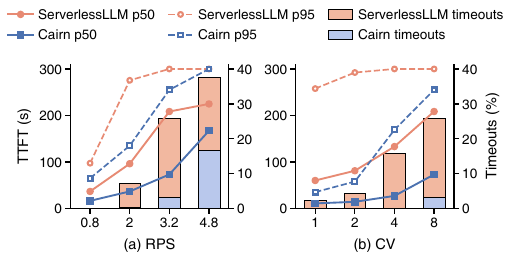}
\vspace{-12pt}
\caption{Simulated TTFT (median and 95th percentile, lines) and timeouts (bars, right axis) for \sllmv{} and \sys{} at keep-alive zero, on a cluster of 16 servers with 4 GPUs each, as the request rate grows at CV~8 (a) and as burstiness grows at 3.2 requests per second (b). Timed-out requests count as 300~s.}
\label{fig:sim}
\end{figure}

To evaluate whether the benefit of \sys{} holds beyond our two servers, we use a discrete-event simulator based on cold starts measured on our servers. Without being calibrated on it, the simulator reproduces the request stream of \S\ref{sec:overlapping}. It overestimates mean TTFT by 11--13\% for \sllmv{} and for \sys{} with either kind of replacement, and it predicts that zygote replacements reduce mean TTFT by 67.2\% compared with \sllmv{}, against 66.5\% measured.

We simulate 16 servers with 4 GPUs each and 160 deployments, copies of our six models treated as distinct models, which share one pool in \sys{}. Requests follow our request-stream benchmark (\S\ref{sec:setup}), and 3.2 requests per second matches the load per GPU on our servers. Decoding takes the ShareGPT output length times the measured time per token. Both systems release a GPU as soon as a request finishes, a request times out if its first token takes more than 300~s, and results are averaged over three seeds.

Figure~\ref{fig:sim}a varies the request rate at CV~8. \sys{} achieves 2.3--3.6$\times$ lower median TTFT than \sllmv{} up to 3.2 requests per second. At 3.2 requests per second, \sys{} reduces the median by 3.6$\times$ and timeouts from 15.6\% to 2.1\%. The gap narrows only at 4.8 requests per second, where both systems saturate the GPUs and \sys{} reduces the median by only 26\%. Figure~\ref{fig:sim}b varies the CV at 3.2 requests per second. \sys{} outperforms \sllmv{} by 3.6--5.2$\times$ in median TTFT at every CV and has no timeouts up to CV~4. In \sys{}, 98\% of instance starts read their weights from the pool, whereas \sllmv{} finds them in DRAM for only 53--84\% and reads the rest from SSD.

\subsection{Overheads}
\label{sec:costs}
An idle warm container with its zygote holds 2.9~GB of host memory (1.8~GB proportional set size, as the zygote shares pages with its children), a 264~MiB CUDA context and 2.5\% of one CPU. Pool entries are written only once per model and configuration. Writing through lengthens that first load from NVMe by 22--23\% (3.85~s compared with 3.12~s for 7B), because writing the pool (about 4~GB/s) is slower than reading the disk.

\section{Conclusion}
A cold start repeats two kinds of work: work whose result is the same on every server, and setup that does not depend on the model. \sys{} removes the first with a cluster-shared pool, from which every server reads a model's state instead of producing it again, and the second with model-agnostic warm containers. Across six models, these cut cold starts to 11.0--21.0~s, 4.1--5.2$\times$ faster, on servers that hold no checkpoint of their own. We will release the source code of \sys{} upon publication.

\bibliographystyle{ACM-Reference-Format}
\bibliography{refs}

@inproceedings{serverlessllm,
  author    = {Fu, Yao and Xue, Leyang and Huang, Yeqi and Brabete, Andrei-Octavian and Ustiugov, Dmitrii and Patel, Yuvraj and Mai, Luo},
  title     = {{ServerlessLLM}: Low-Latency Serverless Inference for Large Language Models},
  booktitle = {18th USENIX Symposium on Operating Systems Design and Implementation (OSDI 24)},
  pages     = {135--153},
  year      = {2024}
}

@inproceedings{blitzscale,
  author    = {Zhang, Dingyan and Wang, Haotian and Liu, Yang and Wei, Xingda and Shan, Yizhou and Chen, Rong and Chen, Haibo},
  title     = {{BlitzScale}: Fast and Live Large Model Autoscaling with {O(1)} Host Caching},
  booktitle = {19th USENIX Symposium on Operating Systems Design and Implementation (OSDI 25)},
  pages     = {275--293},
  year      = {2025}
}

@inproceedings{hydraserve,
  author    = {Lou, Chiheng and Qi, Sheng and Jin, Chao and Nie, Dapeng and Yang, Haoran and Ding, Yu and Liu, Xuanzhe and Jin, Xin},
  title     = {{HydraServe}: Minimizing Cold Start Latency for Serverless {LLM} Serving in Public Clouds},
  booktitle = {23rd USENIX Symposium on Networked Systems Design and Implementation (NSDI 26)},
  pages     = {415--430},
  year      = {2026}
}

@inproceedings{faascale,
  author    = {Yu, Minchen and Yang, Rui and Jia, Chaobo and Su, Zhaoyuan and Yao, Sheng and Lan, Tingfeng and Yang, Yuchen and Wang, Zirui and Cheng, Yue and Wang, Wei and Wang, Ao and Chen, Ruichuan},
  title     = {{FaaScale}: Unlocking Fast {LLM} Scaling for Serverless Inference},
  booktitle = {Proceedings of Machine Learning and Systems (MLSys 26)},
  volume    = {8},
  pages     = {353--366},
  year      = {2026}
}

@inproceedings{warmserve,
  author    = {Lou, Chiheng and Qi, Sheng and Kang, Rui and Zhang, Yong and Sun, Chen and Wang, Pengcheng and Liu, Xuanzhe and Jin, Xin},
  title     = {{WarmServe}: Enabling One-for-Many {GPU} Prewarming for Multi-{LLM} Serving},
  booktitle = {Proceedings of the 43rd International Conference on Machine Learning (ICML 26)},
  series    = {Proceedings of Machine Learning Research},
  volume    = {306},
  year      = {2026}
}

@inproceedings{deepserve,
  author    = {Hu, Junhao and Xu, Jiang and Liu, Zhixia and He, Yulong and Chen, Yuetao and Xu, Hao and Liu, Jiang and Meng, Jie and Zhang, Baoquan and Wan, Shining and Dan, Gengyuan and Dong, Zhiyu and Ren, Zhihao and Liu, Changhong and Xie, Tao and Lin, Dayun and Zhang, Qin and Yu, Yue and Feng, Hao and Chen, Xusheng and Shan, Yizhou},
  title     = {{DeepServe}: Serverless Large Language Model Serving at Scale},
  booktitle = {2025 USENIX Annual Technical Conference (USENIX ATC 25)},
  pages     = {57--72},
  year      = {2025}
}

@inproceedings{medusa,
  author    = {Zeng, Shaoxun and Xie, Minhui and Gao, Shiwei and Chen, Youmin and Lu, Youyou},
  title     = {Medusa: Accelerating Serverless {LLM} Inference with Materialization},
  booktitle = {Proceedings of the 30th ACM International Conference on Architectural Support for Programming Languages and Operating Systems (ASPLOS 25), Volume 1},
  pages     = {653--668},
  year      = {2025}
}

@inproceedings{breakingtheice,
  author    = {Kabakibo, Huzaifa Shaaban and Trivedi, Animesh and Wang, Lin},
  title     = {Breaking the Ice: Analyzing Cold Start Latency in {vLLM}},
  booktitle = {Proceedings of Machine Learning and Systems (MLSys 26)},
  year      = {2026}
}

@inproceedings{vllm,
  author    = {Kwon, Woosuk and Li, Zhuohan and Zhuang, Siyuan and Sheng, Ying and Zheng, Lianmin and Yu, Cody Hao and Gonzalez, Joseph E. and Zhang, Hao and Stoica, Ion},
  title     = {Efficient Memory Management for Large Language Model Serving with {PagedAttention}},
  booktitle = {Proceedings of the 29th ACM Symposium on Operating Systems Principles (SOSP 23)},
  pages     = {611--626},
  year      = {2023}
}

@inproceedings{pytorch2,
  author    = {Ansel, Jason and Yang, Edward and He, Horace and Gimelshein, Natalia and Jain, Animesh and Voznesensky, Michael and Bao, Bin and Bell, Peter and Berard, David and Burovski, Evgeni and Chauhan, Geeta and Chourdia, Anjali and Constable, Will and Desmaison, Alban and DeVito, Zachary and Ellison, Elias and Feng, Will and Gong, Jiong and Gschwind, Michael and Hirsh, Brian and Huang, Sherlock and Kalambarkar, Kshiteej and Kirsch, Laurent and Lazos, Michael and Lezcano, Mario and Liang, Yanbo and Liang, Jason and Lu, Yinghai and Luk, C. K. and Maher, Bert and Pan, Yunjie and Puhrsch, Christian and Reso, Matthias and Saroufim, Mark and Siraichi, Marcos Yukio and Suk, Helen and Suo, Michael and Tillet, Phil and Wang, Eikan and Wang, Xiaodong and Wen, William and Zhang, Shunting and Zhao, Xu and Zhou, Keren and Zou, Richard and Mathews, Ajit and Chanan, Gregory and Wu, Peng and Chintala, Soumith},
  title     = {{PyTorch} 2: Faster Machine Learning Through Dynamic {Python} Bytecode Transformation and Graph Compilation},
  booktitle = {Proceedings of the 29th ACM International Conference on Architectural Support for Programming Languages and Operating Systems (ASPLOS 24), Volume 2},
  pages     = {929--947},
  year      = {2024}
}

@inproceedings{catalyzer,
  author    = {Du, Dong and Yu, Tianyi and Xia, Yubin and Zang, Binyu and Yan, Guanglu and Qin, Chenggang and Wu, Qixuan and Chen, Haibo},
  title     = {Catalyzer: Sub-millisecond Startup for Serverless Computing with Initialization-less Booting},
  booktitle = {Proceedings of the 25th International Conference on Architectural Support for Programming Languages and Operating Systems (ASPLOS 20)},
  pages     = {467--481},
  year      = {2020}
}

@inproceedings{sock,
  author    = {Oakes, Edward and Yang, Leon and Zhou, Dennis and Houck, Kevin and Harter, Tyler and Arpaci-Dusseau, Andrea and Arpaci-Dusseau, Remzi},
  title     = {{SOCK}: Rapid Task Provisioning with Serverless-Optimized Containers},
  booktitle = {2018 USENIX Annual Technical Conference (USENIX ATC 18)},
  pages     = {57--70},
  year      = {2018}
}

@inproceedings{firecracker,
  author    = {Agache, Alexandru and Brooker, Marc and Iordache, Alexandra and Liguori, Anthony and Neugebauer, Rolf and Piwonka, Phil and Popa, Diana-Maria},
  title     = {Firecracker: Lightweight Virtualization for Serverless Applications},
  booktitle = {17th USENIX Symposium on Networked Systems Design and Implementation (NSDI 20)},
  pages     = {419--434},
  year      = {2020}
}

@inproceedings{faasnap,
  author    = {Ao, Lixiang and Porter, George and Voelker, Geoffrey M.},
  title     = {{FaaSnap}: {FaaS} Made Fast Using Snapshot-Based {VMs}},
  booktitle = {Proceedings of the 17th European Conference on Computer Systems (EuroSys 22)},
  pages     = {730--746},
  year      = {2022}
}

@inproceedings{beluga,
  author    = {Yang, Xinjun and Hu, Qingda and Li, Junru and Li, Feifei and Zhu, Yicong and Zhou, Yuqi and Lin, Qiuru and Dai, Jian and Kong, Yang and Zhang, Jiayu and Xu, Guoqiang and Liu, Qiang},
  title     = {Beluga: A {CXL}-Based Memory Architecture for Scalable and Efficient {LLM} {KVCache} Management},
  booktitle = {Proceedings of the 2026 International Conference on Management of Data (SIGMOD 26)},
  year      = {2026}
}

@inproceedings{pond,
  author    = {Li, Huaicheng and Berger, Daniel S. and Hsu, Lisa and Ernst, Daniel and Zardoshti, Pantea and Novakovic, Stanko and Shah, Monish and Rajadnya, Samir and Lee, Scott and Agarwal, Ishwar and Hill, Mark D. and Fontoura, Marcus and Bianchini, Ricardo},
  title     = {Pond: {CXL}-Based Memory Pooling Systems for Cloud Platforms},
  booktitle = {Proceedings of the 28th ACM International Conference on Architectural Support for Programming Languages and Operating Systems (ASPLOS 23), Volume 2},
  pages     = {574--587},
  year      = {2023}
}

@inproceedings{tpp,
  author    = {Maruf, Hasan Al and Wang, Hao and Dhanotia, Abhishek and Weiner, Johannes and Agarwal, Niket and Bhattacharya, Pallab and Petersen, Chris and Chowdhury, Mosharaf and Kanaujia, Shobhit and Chauhan, Prakash},
  title     = {{TPP}: Transparent Page Placement for {CXL}-Enabled Tiered-Memory},
  booktitle = {Proceedings of the 28th ACM International Conference on Architectural Support for Programming Languages and Operating Systems (ASPLOS 23), Volume 3},
  pages     = {742--755},
  year      = {2023}
}

@inproceedings{demystifyingcxl,
  author    = {Sun, Yan and Yuan, Yifan and Yu, Zeduo and Kuper, Reese and Song, Chihun and Huang, Jinghan and Ji, Houxiang and Agarwal, Siddharth and Lou, Jiaqi and Jeong, Ipoom and Wang, Ren and Ahn, Jung Ho and Xu, Tianyin and Kim, Nam Sung},
  title     = {Demystifying {CXL} Memory with Genuine {CXL}-Ready Systems and Devices},
  booktitle = {Proceedings of the 56th Annual IEEE/ACM International Symposium on Microarchitecture (MICRO 23)},
  pages     = {105--121},
  year      = {2023}
}

@misc{safetensors,
  author    = {{Hugging Face}},
  title     = {Safetensors: {ML} Safer for All},
  howpublished = {\url{https://github.com/huggingface/safetensors}},
  year      = {2025}
}

@inproceedings{azurefunctions,
  author    = {Mohammad Shahrad and Rodrigo Fonseca and {\'I}{\~n}igo Goiri and Gohar Chaudhry and Paul Batum and Jason Cooke and Eduardo Laureano and Colby Tresness and Mark Russinovich and Ricardo Bianchini},
  title     = {Serverless in the Wild: Characterizing and Optimizing the Serverless Workload at a Large Cloud Provider},
  booktitle = {2020 USENIX Annual Technical Conference (USENIX ATC 20)},
  pages     = {205--218},
  year      = {2020}
}

@inproceedings{alpaserve,
  author    = {Zhuohan Li and Lianmin Zheng and Yinmin Zhong and Vincent Liu and Ying Sheng and Xin Jin and Yanping Huang and Zhifeng Chen and Hao Zhang and Joseph E. Gonzalez and Ion Stoica},
  title     = {{AlpaServe}: Statistical Multiplexing with Model Parallelism for Deep Learning Serving},
  booktitle = {17th USENIX Symposium on Operating Systems Design and Implementation (OSDI 23)},
  pages     = {663--679},
  year      = {2023}
}

@misc{sharegpt,
  author       = {{ShareGPT}},
  title        = {{ShareGPT} Vicuna Unfiltered Dataset ({ShareGPT\_V3\_unfiltered\_cleaned\_split})},
  howpublished = {\url{https://huggingface.co/datasets/anon8231489123/ShareGPT_Vicuna_unfiltered}},
  year         = {2023}
}

@misc{bedrock,
  author       = {{Amazon Web Services}},
  title        = {Amazon {Bedrock} Custom Model Import},
  howpublished = {\url{https://aws.amazon.com/bedrock/custom-model-import/}},
  year         = {2025}
}

@misc{hfendpoints,
  author       = {{Hugging Face}},
  title        = {Inference Endpoints},
  howpublished = {\url{https://huggingface.co/docs/inference-endpoints}},
  year         = {2025}
}

@misc{predibase,
  author       = {{Predibase}},
  title        = {Introducing the First Purely Serverless Solution for Fine-Tuned {LLMs}},
  howpublished = {\url{https://predibase.com/blog/introducing-the-first-purely-serverless-solution-for-fine-tuned-llms}},
  year         = {2024}
}

@inproceedings{sglang,
  author    = {Zheng, Lianmin and Yin, Liangsheng and Xie, Zhiqiang and Sun, Chuyue and Huang, Jeff and Yu, Cody Hao and Cao, Shiyi and Kozyrakis, Christos and Stoica, Ion and Gonzalez, Joseph E. and Barrett, Clark and Sheng, Ying},
  title     = {{SGLang}: Efficient Execution of Structured Language Model Programs},
  booktitle = {Advances in Neural Information Processing Systems (NeurIPS)},
  year      = {2024}
}

@inproceedings{ray,
  author    = {Moritz, Philipp and Nishihara, Robert and Wang, Stephanie and Tumanov, Alexey and Liaw, Richard and Liang, Eric and Elibol, Melih and Yang, Zongheng and Paul, William and Jordan, Michael I. and Stoica, Ion},
  title     = {Ray: A Distributed Framework for Emerging {AI} Applications},
  booktitle = {13th USENIX Symposium on Operating Systems Design and Implementation (OSDI 18)},
  pages     = {561--577},
  year      = {2018}
}

@misc{trtllm,
  author       = {{NVIDIA}},
  title        = {{TensorRT-LLM}: A {TensorRT} Toolbox for Optimized Large Language Model Inference},
  howpublished = {\url{https://github.com/NVIDIA/TensorRT-LLM}},
  year         = {2023}
}

@inproceedings{triton,
  author    = {Tillet, Philippe and Kung, H. T. and Cox, David},
  title     = {Triton: An Intermediate Language and Compiler for Tiled Neural Network Computations},
  booktitle = {Proceedings of the 3rd ACM SIGPLAN International Workshop on Machine Learning and Programming Languages (MAPL)},
  pages     = {10--19},
  year      = {2019}
}

@article{opt,
  author    = {Zhang, Susan and Roller, Stephen and Goyal, Naman and Artetxe, Mikel and Chen, Moya and Chen, Shuohui and Dewan, Christopher and Diab, Mona and Li, Xian and Lin, Xi Victoria and Mihaylov, Todor and Ott, Myle and Shleifer, Sam and Shuster, Kurt and Simig, Daniel and Koura, Punit Singh and Sridhar, Anjali and Wang, Tianlu and Zettlemoyer, Luke},
  title   = {{OPT}: Open Pre-trained Transformer Language Models},
  journal = {arXiv preprint arXiv:2205.01068},
  year    = {2022}
}

@article{llama3,
  author    = {{Llama Team, AI @ Meta}},
  title   = {The {Llama} 3 Herd of Models},
  journal = {arXiv preprint arXiv:2407.21783},
  year    = {2024}
}

@article{qwen25,
  author    = {{Qwen} and Yang, An and Yang, Baosong and Zhang, Beichen and Hui, Binyuan and Zheng, Bo and Yu, Bowen and Li, Chengyuan and Liu, Dayiheng and Huang, Fei and Wei, Haoran and Lin, Huan and Yang, Jian and Tu, Jianhong and Zhang, Jianwei and Yang, Jianxin and Yang, Jiaxi and Zhou, Jingren and Lin, Junyang and Dang, Kai and Lu, Keming and Bao, Keqin and Yang, Kexin and Yu, Le and Li, Mei and Xue, Mingfeng and Zhang, Pei and Zhu, Qin and Men, Rui and Lin, Runji and Li, Tianhao and Tang, Tianyi and Xia, Tingyu and Ren, Xingzhang and Ren, Xuancheng and Fan, Yang and Su, Yang and Zhang, Yichang and Wan, Yu and Liu, Yuqiong and Cui, Zeyu and Zhang, Zhenru and Qiu, Zihan},
  title   = {{Qwen2.5} Technical Report},
  journal = {arXiv preprint arXiv:2412.15115},
  year    = {2024}
}

\end{document}